\documentclass[intlimits,twoside,a4paper]{article}

\usepackage{amsmath,amssymb}
\usepackage{graphicx}

\usepackage[T2A]{fontenc}
\usepackage[cp1251]{inputenc}

\usepackage[eqsecnum]{cmpj2}
\issue{2013}{16}{3}{33706}
\doinumber{10.5488/CMP.16.33706}

\title[Temperature dependence of plasmon resonances in spheroidal metal nanoparticles]%
{Temperature dependence of plasmon resonances in spheroidal metal nanoparticles%
}
\author[N.I. Grigorchuk]{N.I. Grigorchuk}
\address{
Bogolyubov Institute for
Theoretical Physics of the National Academy of Sciences of Ukraine, \\
 14--b Metrologichna St., 03680 Kyiv, Ukraine
}

\date{Received May 30, 2013, in final form July 1, 2013}
\authorcopyright{N.I. Grigorchuk, 2013}

\begin{document}

\maketitle

\begin{abstract}
The effect of the electron temperature on both the light absorption and the scattering
by metal nanoparticles (MNs) with excitation of the surface plasmon electron vibrations
is studied in the framework of the kinetic theory. The formulae for  electroconductivity
and polarizability tensors are derived for finite temperatures of an electron gas.
The electrical conductivity and the halfwidth of the surface plasmon resonance are studied
in detail for a spherical MN. Depending on the size of MN, the efficiencies of light
absorption and scattering with the temperature change are investigated. It is found, in particular,
that the absorption efficiency can both increase and decrease with a temperature drop.  The derived formulas make it possible to analytically calculate  various optical
and transport phenomena for MNs of any spheroidal shape embedded in any dielectric media.
, {65.80.-g}, {73.23.-b}, {68.49.Jk}, {52.25.Os}
\end{abstract}

\section{Introduction}
\label{intro}
When the metal nanoparticle is illuminated with a laser beam by frequency which coincides
with the frequency of collective electron oscillations in the MN, the surface plasmon resonance
(SPR) is excited. The frequency and the width of SPR depends on the size, morphology, spatial
orientation of MN and on the dielectric environment~\cite{BH,KV}. Resonance light scatterers are
employed in various applications ranging from surface enhanced Raman scattering~\cite{Nie},
near-field scanning optical microscopes~\cite{SD}, to bio-chemical imaging \cite{HRL}, surface
enhanced fluorescence \cite{FG}, subwavelength optical waveguides \cite{JSB}, cancer therapy~\cite{HJE}~etc.

The temperature effect on the optical and transport properties of the metal nanoparticles
is very important for pure and applied science of nanoparticles~\cite{GVH}.
Since a MN absorbs laser energy in a thin-surface layer, rapid local heating can occur at the surface.
Thus, for a detailed analysis of the laser-light absorption or reflection at MN surfaces,
the properties of MNs at electron temperatures must be studied.
The temperature dependence of the SP resonance is crucial due to a number of recent applications
of noble MNs in computer chips~\cite{CWB}, thermally assisted magnetic recording~\cite{CPI}, thermal
cancer treatment~\cite{LGD}, catalysis and nanostructure growth~\cite{CBG}. The use of nano-objects
as temperature sensors~\cite{LAT} and thermometers~\cite{YYL} is quite promising due to their small sizes
and short thermal relaxation time.

Previous calculations of the effect of the temperature on different properties
of the MNs were performed in the zero-temperature limit~\cite{KTT,GT} or
in the interval $15\div1000$~K only~\cite{GVH,Kr,BMC,PBL,LS,FSK_a,FSK_b,FSK_c,FSK_d,WWI,EHD,LPG,BHR,YBG}.
 While the optical properties of both the bulk metals~\cite{RCW,WKL,RMV,GSL,JPA} and the MNs in the
 low-temperature regime are now well studied, the understanding of the effects of electron temperature
 (typically of the order of $10^4$~K) on the plasmon  modes has remained a challenge and is not so well
 understood as at the $T=0$. Such an out-of-equilibrium situation can easily be induced in metallic
nanoparticles~\cite{MM} using ultrashort optical pulses. At time $t=0$, only the electrons
are excited by the laser pulse; then the excited electrons would decay through light radiation, phonon
excitation, and/or electron-electron scattering. In the most commonly used approach, the two-temperature
model~\cite{HBH,GSL,KLT}, assumes that both electronic and ionic degrees of freedom are in
thermal equilibrium conditions individually, but each degree of freedom has its own temperature.
The energy exchange between the electrons and the lattice in this model leads to a time-dependent
electronic temperature $T_\textrm{e}(t)$.

The effect of electron temperature was not studied in detail so far because a broad temperature
interval requires the use of materials with high thermal stability. In \cite{FSK_a,FSK_b,FSK_c,FSK_d,YBG},
an improved Drude model was used where the electron concentration and/or a plasma frequency are directly
dependent on the electron temperature. Only a few works have addressed
this problem so far~\cite{PHK,RGK}.

In this work, we focus on optical properties of metal nanoparticles at relatively high electron
temperatures, whe\-reas the ion temperature, $T_\textrm{i}$, is supposed to be much smaller~\cite{MM}.
Different behavior with temperature of optical lines for MNs with different radii
and shapes is a particular motivation for our present study.
The screening coming from the surrounding matrix is taken into account through a constant
dielectric function, $\epsilon_\textrm{m}$.

To calculate the effect of the electron temperature on the optical and transport characteristics
of MNs, we  use the kinetic theory method which accounts for the electron scattering on the MN
boundary. It is shown that the efficiency of absorption with temperature strongly depends on the
size of MN. It is predicted that the effect of temperature on the MNs with different radii
can substantially modify their optical properties.
We present our theoretical results on the temperature dependence of the conductivity and
polarizability tensors related to the MNs having a spheroidal shape.

The rest of the paper is organized as follows. The Boltzmann equation approach to the problem
is presented in section~2. Section~3 contains the study of conductivity tensor in the spheroidal
MNs at finite temperatures. In section~4, we consider the effect of temperature on the polarizability
of MN. Section~5 is devoted to the study of light absorption and scattering crossection by MNs.
The discussion of the obtained results is available in section~6, while section~7 contains the conclusions.

\section{Boltzmann equation}
\label{sec:1}

To account for the temperature effect on the conductivity and polarization
properties of the MNs, we use the Boltzmann equations approach.
The advantage of this approach is that the obtained results can be applied
not only to MNs with a spherical shape, but to strongly anisotropic spheroidal
(needle-like or disk-shaped) MNs. Thus, it permits to study the effect
of the particle shape on the physical values measured. Second, the Boltzmann
equation method enables us to investigate the MNs having sizes smaller than
the electron mean free path $l$. But in the case of sizes less than $l$,
it provides the same results as known from other approaches.

There exists a lower limit of applicability of Boltzmann method
in a small radius limit when the particle size is comparable to the
de Broglie wavelength of the electron, and the quantization of the electron
spectrum starts to play an essential role~\cite{YB}. Practically,
it is around a radius greater or less than 2~nm.

Let us consider a single ellipsoidal metal MN with semiaxes $a$, $b$, $c$ that
is irradiated by an electromagnetic (EM) wave whose electric field is given as
\begin{equation}
 \label{eq 4}
  {\bf E} = {\bf E}_0 \exp[\ri({\bf k}\cdot{\bf r}-\omega t)].
   \end{equation}
%1
Here, ${\bf E}_0$ is the amplitude of an electric field of a pump laser, $\omega$ is its frequency,
${\bf k}$ is the wave vector, and ${\bf r}$ and $t$ describe the spatial coordinates and time.

We restrict ourselves to the case of Rayleigh scattering, where the electromagnetic wavelength (from a pump
laser) $\lambda\sim c/\omega$ is much larger than the diameter of the nanoparticle $d$ $(= \textrm{max}\{a,b,c\})$.
Then, the electromagnetic field around the MN can be considered as homogeneous. Placing the
coordinate origin in the center of the particle, the above-mentioned assumption is written as follows:
\begin{equation}
 \label{eq 5}
  kr\ll 1.
   \end{equation}
%2

The inequality (\ref{eq 5}) implies that the ${\bf E}$ field of the electromagnetic
wave can be considered to be spatially uniform on scales of the order of a particle
size such that all the conduction electrons move in-phase producing only dipole-type oscillations. The amplitude of such a field is linked to ${\bf E}_0$ by the relation~\cite{LL}
\begin{equation}
 \label{eq if}
  E_0^{(j)}(0,\omega)/E_\textrm{in}^{(j)}(\omega) = 1+L_j [\epsilon(\omega)/\epsilon_\textrm{m}-1],
   \end{equation}
%3
where $\epsilon(\omega)$ is the dielectric permittivity of the MN, $\epsilon_\textrm{m}$
is the dielectric constant of the adjacent medium (i.e., solvent), and $L_j$ are
depolarization factors in the $j$-th direction (in the principal axes of an ellipsoid).

The electric field $E_\textrm{in}$ gives rise to high-frequency current inside the MN.
To obtain the average density of this current over the MN, it is necessary first
of all to calculate the electron velocity distribution function. The field $E_\textrm{in}$
has an effect on the equilibrium electron velocity distribution and thus determines
the appearance of a nonequilibrium addition $f_1({\bf r, v},t)$ to the Fermi distribution
function
$$
f_0(\varepsilon) = \frac{1}{\exp{\left[\frac{\varepsilon-\mu}{k_\textrm{B} T}\right]}+1}\,.
$$
Here, $\mu$ is the chemical potential, $k_\textrm{B}$ is Boltzmann's constant,
$\varepsilon=m\upsilon^2/2 $ denotes the kinetic energy of an electron, $\upsilon=|{\bf v}|$
refers to the electron velocity, and $m$ is the electron mass.
As is well known~\cite{CR_a,CR_b}, the equilibrium function $f_0(\varepsilon)$ does not give any input
to the current. Accounting for both the time dependence of equation~(\ref{eq 4}) and the inequality
(\ref{eq 5}), the total distribution function of electrons can be represented as follows:
\begin{equation}
 \label{eq 7}
  f({\bf r,v},t) = f_0(\varepsilon)+f_1({\bf r,v},t)\equiv f_0(\varepsilon)+f_1({\bf r,v})\,\re^{\ri\omega t}.
   \end{equation}
%4
We seek the function $f_1({\bf r, v})$ as a solution to the linearized Boltzmann's equation
\begin{equation}
 \label{eq BE}
  (\nu-\ri\omega) f_1({\bf r, v})+{\bf v}\frac{\partial f_1({\bf r, v})}{\partial
   {\bf r}}+e{\bf E}_\textrm{in}{\bf v}\frac{\partial f_0(\varepsilon)}{\partial\varepsilon} =0,
    \end{equation}
%5
where $e$ is the electron charge. In equation~(\ref{eq BE}) we have assumed that the collision
integral $$(\partial f_1/\partial t)_\textrm{col} = -f_1/\tau$$ is evaluated in the relaxation
time approximation ($\tau=1/\nu$, $\nu$ refers to electron collision frequency).
Strictly speaking, $\nu\equiv\nu(T)$ is a temperature dependent value which can be presented as follows:
\begin{equation}
 \label{eq nute}
  \nu(T) = \frac{n_\textrm{e} e^2}{m} \frac{K}{\Theta}\left(\frac{T}{\Theta}\right)^5
   \int^{\Theta/T}_0 \frac{z^5 e^z}{(e^z-1)^2}\, \rd z,
    \end{equation}
%6
where $n_\textrm{e}$ is the electron concentration, $\Theta$ is the Debye temperature,
and $K$ combines together the factors depending on the details of the Fermi surface
geometry and scattering matrix elements.

For simplicity, we also assume that the vortex electric field induced by magnetic
component of the external EM field gives a comparatively small input at
the plasmon resonance frequencies and can be neglected in equation~(\ref{eq BE}).

What is more, the function $f_1({\bf r,v})$ ought to satisfy the boundary conditions as well.
These conditions may be chosen from the character of electron reflection from the inner walls of the MN.
We adopt, as is usually done, the assumption of diffusive electron scattering by the boundary of MN.
Then, the boundary conditions can be presented in the form
\begin{equation}
 \label{eq bc}
  f({\bf r,v})|_S = 0, \qquad {\bf v}_{\mathrm{n}}<0,
   \end{equation}
%7
where ${\bf v}_{\mathrm{n}}$ is the velocity of the component normal to the particle surface.

Along with diffusive scattering, the mirror boundary conditions at the nanoparticle
surface were examined in the literature for electron scattering (see, e.g., \cite{CR_a,CR_b}).
In this case, each electron is reflected from the surface at the same angle
at which it falls to the surface. In diffuse reflection, the electron is reflected
from the surface at any angle. For the mirror mechanism to be dominant,
the surface must be perfectly smooth in the atomic scale because the degree of
reflectivity of the boundary essentially depends on its smoothness.
Practically, for a nonplanar border such smoothness is extremely difficult to achieve.
As was shown~\cite{BKY}, the mirror boundary conditions give a small
correction to the results obtained with the account of only the diffusive electron
reflections. Therefore, we choose more realistic boundary conditions given by equation~(\ref{eq bc}).

The boundary conditions (\ref{eq bc}) in the case of an ellipsoidal MN, broadly
speaking, depend on the angles, which complicates the solution of equation~(\ref{eq BE}).

It is rather easy to solve (\ref{eq BE}) and to satisfy the boundary conditions
of (\ref{eq bc}) if one passes to a transformed coordinate system, where an ellipsoid
with semiaxes $a, b, c$ (along the $x$, $y$, and $z$ directions, respectively) transforms
into a sphere of radius $R$ with the same volume:
\begin{equation}
 \label{eq x}
  x_j = \frac{d_j}{R} x'_j\,, \qquad R =(a b c)^{1/3}\,,
   \end{equation}
%8
with $j = 1, \ 2, \ 3$, and $d_1=a$, $d_2=b$, $d_3=c$.
A similar transformation should be made for the electron velocities:
\begin{equation}
 \label{eq v}
  \upsilon_j = \frac{d_j}{R} \upsilon'_j\,.
   \end{equation}
%9

Equation~(\ref{eq BE}) and the boundary conditions (\ref{eq bc}) in
transformed coordinate and velocity systems can be rewritten as follows:
\begin{equation}
 \label{eq bet}
  (\nu-\ri\omega) f_1({\bf r', v'})+{\bf v'}\frac{\partial f_1({\bf r', v'})}{\partial
   {\bf r'}}+ e {\bf E}_\textrm{in}{\bf v'}\frac{\partial f_0(\varepsilon)}{\partial\varepsilon} = 0,
    \end{equation}
%10
\begin{equation}
 \label{eq bct}
  f({\bf r',v'})|_{r'=R} = 0, \qquad {\bf r'}\cdot{\bf v'}<0.
   \end{equation}
%11
The first condition calls for zero equality of the electron distribution function at
the nanoparticle surface, and the second one, ${\bf r'}\cdot{\bf v'}<0$, means that an electron
motion occurs only inside MN, and there is no electron leakage through the NP surface.

Equation~(\ref{eq bet}) presents the partial differential equation with boundary conditions (\ref{eq bct}).
Using the method of characteristics, one can obtain the solution to this equation in the form
\begin{equation}
 \label{eq f1}
  f_1({\bf r',v',}\,t) = -e\frac{\partial f_0}{\partial\varepsilon}
   {\bf v' E}_\textrm{in}\frac{1-\exp[-(\nu-\ri\omega)\;t_\textrm{c}({\bf r',v'})]}{\nu-\ri\omega}\,,
    \end{equation}
%12
where the characteristic $t_\textrm{c}({\bf r',v'})$ can be presented as follows:

\begin{equation}
 \label{eq char}
  t_\textrm{c}({\bf r',v'}) = \frac{1}{{\bf v}^{'2}}\left[\bf r'v'+\sqrt{(R^2-r^{'2})
   \,{\bf v}^{'2}+({\bf r' v'})^2}\right].
   \end{equation}
%13
The radius vector ${\bf R}$ determines the starting position of an electron at the moment $t_\textrm{c}=0$.
The characteristic curve of equation~(\ref{eq char}) depends only on the absolute value
of ${\bf R}$ and does not depend on the direction of ${\bf R}$.

It is reasonable to point out that, in spite of the electric field which remains spatially
uniform inside the MN, the distribution function (\ref{eq f1}) still  depends on
the coordinates due to the requirement to obey the boundary conditions (\ref{eq bct}).
Owing to this dependence, other physical parameters averaged with $f_1({\bf r',v'})$ start
to depend on coordinates too. Since the physical sensing involves only parameters averaged over the
whole MN volume $V$, it is necessary to fulfill the integration over all the coordinates inside the MN.
One can see this exemplified just below.

Using the solution (\ref{eq f1}), one can calculate the density of a high-frequency current induced
by the EM wave inside the MN. Performing the Fourier transformation of equation~(\ref{eq f1}), we obtain
\begin{equation}
 \label{eq st}
  {\bf j}(\omega) = \frac{2e}{V}\left(\frac{m}{2\pi\hbar}\right)^3 \int\!\!\!\int\limits_V\!\!\!\int \rd^3 r'
   \int\!\!\!\int\!\!\!\int \rd^3 \upsilon'\,{\bf v'} f_1({\bf r',v',}\,\omega).
    \end{equation}
%14

\section{Electric-conductivity tensor}
\label{sec:2}

Let us introduce the tensor of electric conductivity $\sigma_{\alpha\beta}(\omega)$
using the relationship
\begin{equation}
 \label{eq ca}
  j_{\alpha}(\omega) = \sum\limits_{\beta=1}^{3}\sigma_{\alpha\beta}(\omega)\,E_\textrm{in}^{(\beta)}(\omega).
   \end{equation}
%15

Then, in accordance with both equations~(\ref{eq f1}) and  (\ref{eq st}),
the components of this tensor can be presented as follows:
\begin{eqnarray}
 \label{eq sg}
  \sigma_{\alpha\beta}(\omega) = - \frac{2e}{V}\left(\frac{m}{2\pi\hbar}\right)^3
   \int\!\!\!\int\limits_V\!\!\!\int\limits \rd^3 r'
    \int\!\!\!\int\!\!\!\int \rd^3\upsilon'\;\upsilon_{\alpha}   \left[e\upsilon_{\beta}\;\frac{\partial f_0}{\partial\varepsilon}
      \;\frac{1-\re^{-(\nu-\ri\omega)t_\textrm{c}(r',\upsilon')}}{\nu-\ri\omega}\right].
       \end{eqnarray}
%16
Electric-conductivity tensor $\sigma_{\alpha\beta}=\sigma_{\beta\alpha}$ is the symmetrical
second-rank tensor. Since the integrand in equation~(\ref{eq sg}) is an odd function for nondiagonal
tensor components $\alpha\neq\beta$, and the integration is conducted over the whole velocity
space [$-\infty$, $\infty$], only diagonal components in this equation have been retained.

If we introduce a rectangular coordinate system and choose the fundamental directions $x$, $y$, $z$ to be coincident with the three principal axes of the ellipsoids, then we get
\begin{displaymath}
 \label{eq mat}
  \mathbf{\sigma}=\left(\begin{array}{ccc}
   \sigma_{xx} & 0 & 0 \\
    0 & \sigma_{yy} & 0 \\
     0 & 0 & \sigma_{zz}
     \end{array}\right).
      \end{displaymath}
The tensor of electric conductivity, as can be seen from equation~(\ref{eq sg}), is a complex value:
\begin{equation}
 \label{eq sc}
  \sigma_{\alpha\beta}(\omega) =  \sigma'_{\alpha\beta}(\omega) + \ri  \sigma''_{\alpha\beta}(\omega).
   \end{equation}
%17

The surface effect on the conducting phenomenon is described in equation~(\ref{eq sg})
by means of a characteristic $t_\textrm{c}({\bf r', v'})$. It accounts for the restrictions
  imposed on the electron movement by nanoparticle surfaces. As one can see from
  equation~(\ref{eq f1}), the value of $t_\textrm{c}$ is of the order of $t'\sim R/\upsilon_\textrm{F}$,
  where $\upsilon_\textrm{F}$ is the Fermi velocity. This implies that the value reciprocal
  to $t_\textrm{c}$ will correspond to the vibration frequency between the particle walls.
  Hence, the inequality $\nu t_\textrm{c}\gg 1$ indicates that the electron collision frequency
  inside the MN bulk would significantly exceed the one for an electron collision
  with the surface of MN. If this inequality is satisfied, one can direct $t_\textrm{c}\rightarrow\infty$.
  Then, the exponent in equation~(\ref{eq sg}) vanishes and finally we arrive at a standard
  expression for an electric-conductivity \cite{LL,BH}
\begin{equation}
 \label{eq sgs}
  \sigma(\omega) = \frac{1}{4\pi}\frac{\omega^2_\textrm{pl}}{\nu-\ri\omega}\,,
   \end{equation}
%18
with
\begin{equation}
 \label{eq op}
  \omega^2_\textrm{pl} = \frac{4 e^2 m^2\upsilon_\textrm{F}^3}{3\pi\hbar^3}\,.
   \end{equation}
%19

To be sure of that, it is necessary to take into account that the energy
derivative of $f_0$ in the zero approximation in the small ratio of
$k_\textrm{B} T/\varepsilon_\textrm{F}$ ($\varepsilon_\textrm{F}$ is the Fermi energy) can be replaced by
\begin{equation}
 \label{eq 20}
  \frac{\partial f_0}{\partial\varepsilon}\approx -\delta(\varepsilon-\varepsilon_\textrm{F}),
   \end{equation}
%20
and passes in equation~(\ref{eq sg}) to the integration over $\upsilon$ in the spherical coordinate system
\[
\int\!\!\!\int\limits_{-\infty}^{\infty}\!\!\!\int\;\rd^3\;
\upsilon\rightarrow\int\limits_{0}^{2\pi}\rd\varphi
\int\limits_0^{\pi}\sin\theta\,\rd\theta\int\limits_0^{\infty}\upsilon^2\,\rd\upsilon,
\]
with the use of the formula
\begin{equation}
 \label{eq 16}
  \int_0^{\infty}\upsilon^4\,\delta(\upsilon^2-\upsilon^2_\textrm{F})
   \, \rd\upsilon = \frac{\upsilon^3_\textrm{F}}{2}\,.
   \end{equation}
%21
We denote with $\varphi$ and $\theta$ the azimuthal and polar angles with
respect to the ellipsoid rotation axis $z$, respectively. Note here that only
diagonal terms with $\upsilon_{\alpha}=\upsilon_{\beta}=\upsilon$ are retained
after integration over all angles. As one can see from equation~(\ref{eq sgs}),
the conductivity becomes a scalar quantity in this approximation.

In the general case of an ellipsoidal-shaped MN, the electric conductivity is the
tensor quantity (\ref{eq sg}). Integrating it over all nanoparticle coordinates
in a solid angle $\rd\Omega=\sin\theta\;\rd\varphi\;\rd\theta$ gives

\begin{equation}
 \label{eq ci}
  \frac{1}{V}\int\!\!\!\int\limits_V\!\!\!\int\limits \rd^3 r'
   \left[1-\re^{-(\nu-\ri\omega)t'({\bf r',\upsilon'})}\right] = \frac{3}{4}\Psi(\omega,\upsilon'),
    \end{equation}
%22
where we use

\begin{equation}
 \label{eq A3}
  \int\limits_0^{\infty} \rd z\, z^3 \,{\rm Erfc}{\left(z+\frac{a}{z}\right)} =
   \frac{1}{4} \left(a+\frac{3}{4}\right)\,\re^{-4a},
\end{equation}
provided that ${\rm Re}\, a > 0$, and

\begin{equation}
 \label{eq A4}
  \int\limits_0^{\infty} \rd z\, z^3 \,{\rm Erfc}{\left(z-\frac{a}{z}\right)} =
   \frac{1}{2} \left(a^2+a+\frac{3}{8}\right).
\end{equation}
The complex $\Psi$ function entering the equation~(\ref{eq ci}) has the form
\begin{equation}
 \label{eq psi}
  \Psi(\omega,\upsilon') = \Phi(\omega,\upsilon')-\frac{4}{q^2}\left(1+\frac{1}{q}\right)\re^{-q},
\end{equation}
%23
with
\begin{equation}
 \label{eq q}
  \Phi(\omega,\upsilon') = \frac{4}{3}-\frac{2}{q}+\frac{4}{q^3}\,,
   \qquad   q \equiv q(\omega,\upsilon') = \frac{2R}{\upsilon'}(\nu-\ri\omega)\,,
\end{equation}
%24
and $\upsilon'$ $(=\varsigma\upsilon)$ is a ``deformed'' electron velocity \cite{GT} with
the coefficient of ``deformation'' $\varsigma_j=R/d_j$. The last summand in equation~(\ref{eq psi})
represents the oscillation part of the $\Psi$ function and the first one refers to its smooth part.

Accounting for equation~(\ref{eq ci}) and only diagonal components in equation~(\ref{eq sg}), there remain
the integrals over all electron velocities. To calculate them, we pass to a spherical coordinate
system with the $z$ axis directed along the rotation axis of the spheroid (as we have done it above).
Then, equation~(\ref{eq sg}) can be rewritten as follows:
\begin{eqnarray}
 \label{eq sgm}
  \sigma_{jj}({\omega})  = \frac{3e^2 m^3}{2(2\pi\hbar)^3}
   \frac{1}{\nu-\ri\omega} \int\limits_0^{2\pi} \rd\varphi\int
    \limits_0^{\pi}\sin\theta\; \rd\theta
    \int\limits_0^{\infty} \upsilon^2 \rd\upsilon\,\upsilon^2_{j}\left( -\frac{\partial f_0}{\partial\varepsilon}\right)\,\Psi(\omega,\upsilon'),
     \end{eqnarray}
%25
where $\upsilon_j$ is the $j$-th component of the electron velocity, $j = x, y, z$, with
\[
\left\{\begin{array}{l}
\upsilon^2_{x\choose y} = \upsilon^2\sin^2\theta\cdot{\cos^2\varphi\choose\sin^2\varphi},\\[2ex]
\upsilon^2_z = \upsilon^2\cos^2\theta,
\end{array}\right.
\]
respectively.

It should be noted that the ``deformed'' electron velocity entering the $\Psi$ function can
be expressed through the electron velocity in a Cartesian coordinate system as follows:
\begin{equation}
 \label{eq vv}
  \upsilon' = \upsilon R \sqrt{\left(\frac{\cos^2\varphi}{a^2}+
   \frac{\sin^2\varphi}{b^2}\right)\sin^2\theta +\frac{\cos^2\theta}{c^2}}\,.
\end{equation}
%26

Let us suppose that the particle in the matrix is modelled as a rotationally symmetric ellipsoid
($a = b\equiv R_{\bot}$, $c\equiv R_{\|}$) with the symmetry axis along the $z$ direction.
The components of an electron velocity parallel ($\upsilon_{\|}$)
and perpendicular ($\upsilon_{\bot}$) to the spheroid revolution axis
\begin{equation}
 \label{eq ups}
  \upsilon_{\|} = \upsilon_z=\upsilon\cos\theta, \qquad \upsilon_{\bot}=
   \sqrt{\upsilon^2_x+\upsilon^2_y}=\upsilon\sin\theta
    \end{equation}
%27
play an important role in this case, and the $\upsilon'$ ceases to depend on the angle $\varphi$.

Let us pass from an integration in equation~(\ref{eq sgm}) over electron
velocities to the integration over electron energies
\begin{equation}
 \label{eq ve}
  \upsilon^4\, \rd\upsilon = \frac{1}{m}\left(\frac{2\varepsilon}{m}\right)^{3/2}\,\rd\varepsilon
   \end{equation}
%28
and take into account the integrals
\begin{equation}
 \label{eq ints}
  \int\limits_0^{2\pi} \cos^2\varphi \; \rd\varphi = \int\limits_0^{2\pi} \sin^2\varphi \;  \rd\varphi = \pi.
\end{equation}
%29
Then, with the use of equations~(\ref{eq sgm}), (\ref{eq ve}), and (\ref{eq ints}),
we obtain for the main components of the complex electric conductivity tensor, namely for
\begin{equation}
 \label{eq ss}
  \sigma_{xx} = \sigma_{yy} \equiv \sigma_{\bot}\,, \qquad  \sigma_{zz} = \sigma_{\|}\,,
\end{equation}
%30
the expression
\begin{eqnarray}
 \label{eq sigt}
  \sigma_{\|\choose\bot}(\omega) = \frac{3e^2}{2\pi^2}
   \frac{\sqrt{2m}}{\hbar^3}\left[\frac{1}{\nu-\ri\omega}
    \int\limits_0^{\pi/2}{\cos\theta'\,\sin^2\theta'\choose\frac{1}{2}\cos^3\theta'}
     \,\rd\theta'\int_0^{\infty}\Psi(\omega,\varepsilon'_{\theta'})
      \left(-\frac{\partial f_0}{\partial\varepsilon}\right)\varepsilon^{3/2}\,\rd\varepsilon\right].
       \end{eqnarray}
%31
We denote with $\theta' (=\theta-\pi/2)$ the angle between the direction of an electron velocity
and an axis perpendicular to the spheroid rotation axes. Here, and below, the upper (lower) symbol
in the parentheses on the left-hand side of equation~(\ref{eq sigt}) corresponds
to the upper (lower) expression in the parentheses on the right-hand side of this equation.

The $\Psi$ function in (\ref{eq sigt}) depends now on both the energy $\varepsilon'$
and the angle $\theta'$ because the parameter $q$ [see (\ref{eq q}) and (\ref{eq vv})]
for a spheroidal particle becomes dependent on the angle $\theta$ and can be determined as
\begin{equation}
 \label{eq ku}
  q = \sqrt{\frac{2m}{\varepsilon}}
   {\left(\nu-\ri\omega\right)}
   {{\left(\frac{\sin^2\theta}{R^2_{\bot}}+
    \frac{\cos^2\theta}{R^2_{\|}}\right)^{-1/2}}}
    \equiv q(\theta,\varepsilon),
     \end{equation}
%32
where $R_{\|}$ and $R_{\bot}$ are the semiaxes of the spheroid.
The electron energy $\varepsilon'$ entering the $\Psi$ function,
becomes dependent on the angle $\theta$ too, and for a spheroid can be presented as follows:
\begin{eqnarray}
 \label{eq vet}
  \varepsilon'_{\theta'} = \varepsilon R^2 \left(\frac{\cos^2\theta'}{R^2_{\bot}}+
   \frac{\sin^2\theta'}{R^2_{\|}}\right),
    \end{eqnarray}
%33
where the equation~(\ref{eq vv}) was used. Since the spheroid semiaxes $R_{\bot}$
and $R_{\|}$ can be easily expressed through the radius
of a sphere $R$ of an equivalent volume
\begin{equation}
 \label{eq r}
  R_{\bot} = R \left(\frac{R_{\bot}}{R_{\|}}\right)^{1/3},
   \qquad R_{\|} = R \left(\frac{R_{\bot}}{R_{\|}}\right)^{-2/3},
    \end{equation}
%34
the energy $\varepsilon'$ does not depend on the particle
radius $R$ but depends only on the spheroid axes ratio.

If we restrict ourselves here to the case of low temperatures,
then the equation~(\ref{eq 20}) can be used, and we find
\begin{equation}
 \label{eq intlt}
  \int_0^{\infty} \Psi(\omega,\varepsilon'_{\theta'})\; \delta(\varepsilon-
   \varepsilon_\textrm{F}) \,\varepsilon^{3/2}\, \rd\varepsilon = \mu_0^{3/2} \Psi(\omega,\mu'_{\theta'})\,,
    \end{equation}
%35
where $\mu_0$ is the chemical potential at zero temperature.
Substituting (\ref{eq intlt}) into equation~(\ref{eq sigt}) and accounting
that the electron concentration at $T=0$ can be presented as
\begin{equation}
 \label{eq n}
  n_0 = \frac{(2m\mu_0)^{3/2}}{3\pi^2\hbar^3}\,,
   \end{equation}
%36
it is easy to check that equation~(\ref{eq sigt}) transforms to the form known from (see, e.g., \cite{G}).

\subsection{Conductivity of a spherical MN}
\label{sec:2.1}

For particles having a spherical shape, the electric conductivity becomes a scalar
quantity, and one can put $R_{\|} = R_{\bot}\equiv R$, $\varepsilon' = \varepsilon$
in  equations~(\ref{eq ku}) and (\ref{eq ve}); then, $q$ and the $\Psi$ function
cease to dependent on the angle $\theta'$,
\begin{equation}
 \label{eq teta}
  \int\limits_0^{\pi/2}{\cos\theta'\,\sin^2\theta'
   \choose\frac{1}{2}\cos^3\theta'} \,\rd\theta' = \frac{1}{3}\,,
    \end{equation}
%37
and equation~(\ref{eq sigt}) reduces to the form
\begin{equation}
 \label{eq ssph}
  \sigma_\textrm{sph}(\omega) = \frac{e^2\sqrt{2m}}{2\pi^2\hbar^3(\nu-\ri\omega)}
    \int_0^{\infty}\Psi(\omega,\varepsilon)
     \left(-\frac{\partial f_0}{\partial\varepsilon}\right)\varepsilon^{3/2}\,\rd\varepsilon.
      \end{equation}
%38

Let us make the following change of variables in equation~(\ref{eq ssph})
\begin{equation}
 \label{eq eta}
  \eta = \frac{\varepsilon-\mu}{k_\textrm{B} T}, \qquad \rd\eta = \frac{\rd\varepsilon}{k_\textrm{B} T}\,.
   \end{equation}
%39
Then
\begin{equation}
 \label{eq dfde}
- \frac{\partial f_0}{\partial\varepsilon} \rd \varepsilon= \frac{\re^{-\eta}}{(1+\re^{-\eta})^2} \rd\eta\,,
   \end{equation}
%40
and the integral in (\ref{eq ssph}) can be presented as
\begin{equation}
 \label{eq intet}
  \int\limits_{-\mu/(k_\textrm{B} T)}^{\infty}\Psi(\omega,\eta)\,(\mu+\eta k_\textrm{B} T)^{3/2}
   \frac{\re^{-\eta}}{\left(1+\re^{-\eta}\right)^2}\,\rd\eta \equiv I,
    \end{equation}
%41
where
\begin{eqnarray}
 \label{eq psi0}
  \Psi(\omega,\eta) &=& \frac{4}{3}-\frac{\sqrt{\frac{2}{m}(\mu+\eta kT)}}{R(\nu-\ri\omega)}+
    \frac{\left[\frac{2}{m}(\mu+\eta kT)\right]^{3/2}}{2R^3(\nu-\ri\omega)^3}
     \nonumber \\&&-\frac{\frac{2}{m}(\mu+\eta kT)}{R^2(\nu-\ri\omega)^2}\left[1+
      \frac{\sqrt{\frac{2}{m}(\mu+\eta kT)}}{2R(\nu-\ri\omega)}\right] \exp\left[-
       \frac{2R(\nu-i\omega)}{\sqrt{\frac{2}{m}(\mu+\eta kT)}}\right],\nonumber\\
        \end{eqnarray}
%42
with~\cite{L}
\begin{equation}
 \label{eq mut}
  \mu\equiv\mu(T)\simeq \mu_0\left[1-\frac{\pi^2}{12}\left(\frac{k_\textrm{B} T}{\mu_0}\right)^2\right].
    \end{equation}
%43
Since the input of the integrand tends to zero as $\eta\rightarrow -\infty$,
the lower limit in this integral can be extended to the $-\infty$.
The derivative $(-\partial f_0/\partial\varepsilon)$ has a maximum at the point $\varepsilon = \mu$. So, it is
conveniently to expand the product of $\Psi\cdot(\mu+\eta k_\textrm{B} T)^{3/2}\equiv\Xi(\eta)$ in the powers of $\eta$
\begin{eqnarray}
 \label{eq_expan1}
  \Xi(\eta)&\simeq& \mu^{3/2}\Psi(0)+\left[\frac{3}{2}k_\textrm{B} T\sqrt{\mu}\Psi(0)+\mu^{3/2}\Psi'(0)\right]\eta
   \nonumber \\&&+ \frac{1}{8\sqrt{\mu}}\left[3(k_\textrm{B} T)^2 \Psi(0)+12k_\textrm{B} T\mu\Psi'(0)+4\mu^2\Psi''(0)\right]
   \eta^2+ 0[\eta]^3,
     \end{eqnarray}
%44
with
\begin{eqnarray}
 \label{eq psi3}
  &&\Psi(0)=\Psi(\omega,\eta)\Big|_{\eta\rightarrow 0}\,, \qquad \Psi'(0) = \frac{\partial}{\partial\eta}
  \Psi(\omega,\eta)\Big|_{\eta\rightarrow 0}\,, \qquad \Psi''(0)= \frac{\partial^2}{\partial\eta^2}
   \Psi(\omega,\eta)\Big|_{\eta\rightarrow 0}.
    \end{eqnarray}
%45
Note, that $\Psi(0)$ at $\upsilon'=\upsilon_\textrm{F}$ coincides with the $\Psi(\omega)$ defined by equation~(\ref{eq psi}).

Then, the integral (\ref{eq intet}) can be presented as the sum of integrals
\begin{eqnarray}
 \label{eq_expan2}
   I &\simeq & \mu^{3/2}\Psi(0)\;I_1 + \left[\frac{3}{2}k_\textrm{B} T
  \sqrt{\mu}\Psi(0)+ \mu^{3/2}\Psi'(0)\right]I_2
   \nonumber \\&&+ \frac{1}{2}\left[\frac{3(k_\textrm{B} T)^2}{4\sqrt{\mu}}
    \Psi(0)+3k_\textrm{B} T\sqrt{\mu}\Psi'(0)+\mu^{3/2}\Psi''(0)\right]I_3\,,
      \end{eqnarray}
%46
where the integrals $I_{j}=\int_{-\infty}^\infty \rd \eta \, {\eta^{j-1}\exp(-\eta)}\big/{\left[1+\exp(-\eta)\right]^2}$ are calculated to be: $I_1=1$, $I_2=0$ and $I_3=\pi^2/6$.

Using $I_{j}$ and (\ref{eq eta}), equation~(\ref{eq ssph}) transforms into
\begin{eqnarray}
 \label{eq sigav}
  \sigma_\textrm{sph}(\omega) &=& \frac{e^2\sqrt{2m\mu}}{2\pi^2\hbar^3(\nu-\ri\omega)}
   \left\{\mu\left[1+\frac{\pi^2}{8}\left(\frac{k_\textrm{B} T}{\mu}\right)^2\right]\Psi(0)%
   +\frac{\pi^2}{6}\left[3k_\textrm{B} T\;\Psi'(0)+\mu\Psi''(0)\right] + 0[\eta]^3 \right\}.
     \end{eqnarray}
%49
Here,
\begin{eqnarray}
 \label{eq psi1}
  \Psi'(0)& =&
   \frac{k_\textrm{B} T}{\mu q_{\mu}}\left[-1+\frac{6}{q_{\mu}^2}-\frac{6}{q_{\mu}}\re^{-q_{\mu}}
    \left(1+\frac{1}{q_{\mu}}+\frac{q_{\mu}}{3}\right)\right],
\\
 \label{eq psi2}
  \Psi''(0)& =&
   \frac{(k_\textrm{B} T)^2}{2\mu^2 q_{\mu}}\left[1 +\frac{6}{q_{\mu}^2} - 4 \re^{-q_{\mu}}
     \left(1+\frac{q_{\mu}}{2}+\frac{3}{2q_{\mu}}+\frac{3}{2 q^2_{\mu}}\right)\right],
      \end{eqnarray}
%51
with
\begin{equation}
 \label{eq qmu}
  q_{\mu} = \frac{2R(\nu-\ri\omega)}{\sqrt{2\mu/m}}\,.
   \end{equation}
%52

Substituting (\ref{eq psi1})--(\ref{eq psi2}) into equation~(\ref{eq sigav}), we obtain
\begin{eqnarray}
 \label{eq sigmT}
  \sigma_\textrm{sph}(\omega,T) &=& \frac{me^2\mu R}{\pi^2\hbar^3q_{\mu}}
   \left[\Psi(0)+\alpha_T \Psi_T(\omega)\right],
     \end{eqnarray}
%53
with
\begin{equation}
 \label{eq alT}
  \alpha_T = \frac{\pi^2}{6}\left(\frac{k_\textrm{B} T}{\mu}\right)^2,
   \end{equation}
%54
\begin{equation}
 \label{eq psT}
  \Psi_T(\omega) = \Phi_T(\omega)-\left[1+\frac{8}{q_{\mu}}+
   \frac{24}{q^2_{\mu}}\left(1+\frac{1}{q_{\mu}}\right)\right]\re^{-q_{\mu}},
    \end{equation}
%55
\begin{equation}
 \label{eq fiT}
  \Phi_T(\omega) = 1-\frac{4}{q_{\mu}}+\frac{24}{q^3_{\mu}}\,.
    \end{equation}
%56

When the temperature $T\rightarrow 0$ in equation~(\ref{eq alT}),
then (\ref{eq sigmT}) is reduced to the form
\begin{equation}
 \label{eq sigms}
  \sigma_\textrm{sph}(\omega) = \frac{e^2\mu_0^{3/2}\sqrt{2m}}{2\pi^2\hbar^3}\frac{\Psi(0)}{\nu-\ri\omega}\,,
   \end{equation}
%57
known from other calculations (see, e.g., \cite{G2}).

The conductivity of MN given by (\ref{eq sigmT}) is a complex value.
Practically, we need to know both the real and the imaginary parts of it.

Further analytical calculations are possible for some particular cases.
In the case of a spherical MN, there are three actual frequencies that
are usually considered: the frequency of an incident electromagnetic field $\omega$,
the collision frequency of electrons in the particle volume $\nu$, and the vibration
frequency between the particle walls $\nu_{\mathrm{s}}=\upsilon_\textrm{F}/(2R)$ (provided the particle size is less than
the electron mean free path). When $\nu>\nu_{\mathrm{s}}$, the mechanism of an electron scattering
in the bulk dominates, and an electron scattering from the particle surface gives
only small corrections of the order of $\nu_{\mathrm{s}}/\nu$. However, a much more interesting case is when
the mechanism of the surface electron scattering dominates, which corresponds to the inequality $\nu<\nu_{\mathrm{s}}$.

Let us suppose that the mean free path of a conduction electron $l$ is much
greater than the particle size. In this case, an electron scattering occurs
mainly from the inner surface of the MN. The electrons oscillate between
the walls of the particle with the frequency $\nu_{\mathrm{s}}$. In the case of $\nu\ll\nu_{\mathrm{s}}$,
to a first approximation, one can formally put $\nu\rightarrow 0$.
If one introduces the designation
\begin{equation}
 \label{eq 55}
  q_i = \frac{\omega}{\nu_{\mathrm{s}}}\,,
    \end{equation}
%58
then for real and imaginary parts of the ratios
$\Psi/(\nu-\ri\omega)$ and $\Psi_T/(\nu-\ri\omega)$ we obtain
\begin{align}
 \label{eq re}
&  {\rm Re}\left[\frac{\Psi}{\nu-\ri\omega}\right]_{\nu\rightarrow 0} =
   \frac{1}{\omega}\left[\frac{2}{q_i}-\frac{4}{q^2_i}
    \sin q_i+\frac{4}{q_i^3}(1-\cos q_i)\right],\\
%     \end{equation}
%59
%\begin{equation}
 \label{eq im}
&  {\rm Im}\left[\frac{\Psi}{\nu-\ri\omega}\right]_{\nu\rightarrow 0} = \frac{4}{\omega}
   \left[\frac{1}{3}+ \frac{1}{q^2_i}\left(\cos q_i-\frac{\sin q_i}{q_i}\right)\right],
    \end{align}
%60
and
\begin{align}
 \label{eq reT}
 &{\rm Re}\left[\frac{\Psi_T}{\nu-\ri\omega}\right]_{\nu\rightarrow 0} =
   \frac{1}{\omega}\left[\left(1-\frac{24}{q^2_i}\right)\sin q_i +\frac{4}{q_i}\left(1+\frac{6}{q^2_i}\right)+
     \frac{8}{q_i}\left(1-\frac{3}{q^2_i}\right)\cos q_i\right],\\
%      \end{eqnarray}
%61
%\begin{eqnarray}
 \label{eq imT}
&  {\rm Im}\left[\frac{\Psi_T}{\nu-\ri\omega}\right]_{\nu\rightarrow 0} =
   \frac{1}{\omega}\left[1-\left(1-\frac{24}{q^2_i}\right)\cos q_i +\frac{8}{q_i}\left(1-\frac{3}{q^2_i}\right)\sin q_i\right].
     \end{align}
%62
Substituting (\ref{eq re})--(\ref{eq imT}) into equation~(\ref{eq sigmT}),
one obtains for the real and imaginary parts of $\sigma$ the following expressions:
\begin{eqnarray}
 \label{eq sig1}
  \sigma'_\textrm{sph}(\omega,T) &=& \frac{3}{4}\frac{e^2 n_\textrm{e}(T)}{m\omega}\left\{
    \frac{2}{q_i}-\frac{4}{q^2_i}\sin q_i+\frac{4}{q_i^3}(1-\cos q_i)\right.\nonumber\\
     && + \left.\alpha_T\left[\frac{4}{q_i}+\frac{24}{q^3_i}+
      \left(1-\frac{24}{q_i^2}\right)\sin q_i+\frac{8}{q_i}
       \left(1-\frac{3}{q_i^2}\right)\cos q_i\right]\right\},\\
%        \end{eqnarray}
%63
%\begin{eqnarray}
 \label{eq sig2}
  \sigma''_\textrm{sph}(\omega,T)& =& \frac{3}{4}\frac{e^2 n_\textrm{e}(T)}{m\omega}\left\{
    \frac{4}{3}+ \frac{4}{q^2_i}\left(\cos q_i-\frac{\sin q_i}{q_i}\right)\right.
     \nonumber \\&&+ \left.\alpha_T\left[1+\frac{8}{q_i}\left(1-\frac{3}{q^2_i}\right)\sin q_i -
      \left(1-\frac{24}{q^2_i}\right)\cos q_i\right]\right\},
       \end{eqnarray}
%64
provided that $\nu\ll\nu_{\mathrm{s}}$. In the presented equations, the electron
concentration depends on the temperature as~\cite{L}
\begin{equation}
 \label{eq_ect}
  n_\textrm{e}(T) = n_0 \left\{1+\frac{\pi^2}{8}\left[\frac{k_\textrm{B} T}{\mu}\right]^2\right\}.
    \end{equation}
%65
If one neglects the oscillation terms in the equation~(\ref{eq sig2}), then we obtain
\begin{equation}
 \label{eq DS}
  \sigma''_\textrm{sph}(\omega,T)\simeq\frac{\omega^2_\textrm{pl}(T)}{4\pi\omega}
   \left(1+\frac{3}{4}\alpha_T\right),
   \end{equation}
%66
with $\omega^2_\textrm{pl}(T)=4\pi e^2 n_\textrm{e}(T)/m$.

At $T\rightarrow 0$, equations~(\ref{eq sig1}) and (\ref{eq sig2})
agree with the expressions known from earlier calculations \cite{G,G2}.

\subsection{Linewidth and ``figure of merit''}
\label{sec:2.2}

The linewidth of the particle-plasmon resonance is controlled by lifetime broadening due to
various decay processes.
Knowing the expressions for conductivity tensor (\ref{eq sigt}) and (\ref{eq sig1}),
the halfwidth of the plasmon line absorption can be calculated simply by means of a relation
\begin{equation}
 \label{eq gam}
  \Gamma_{j}(\omega,T)=\frac{4\pi}
   {\epsilon_{\infty} + n^2(1/L_{j}-1)} {\rm Re}\,\sigma_{jj}(\omega,T),
    \end{equation}
%69
where $n (= \sqrt{\epsilon_\textrm{m}})$ is the refractive index of the surrounding medium
and $\varepsilon_{\infty}\equiv 1+\epsilon_{\rm inter}$ is the high frequency dielectric
constant due to interband and core transitions of the inner electrons in a MN's material.

The ``figure of merit'' (FOM) can be expressed through the $\sigma_{jj}$ as well.
For MNs having a spheroidal shape we found~\cite{G2}
\begin{equation}
 \label{eq fom}
   {\rm FOM}_{\|\choose\bot} = \frac{n \omega_{\rm pl}\left[1/L_{\|\choose\bot}-1\right]}
    {4\pi\sqrt{\varepsilon_{\infty}+\left[1/L_{\|\choose\bot}-1\right]n^2}}
     \sigma'^{-1}_{\|\choose\bot}(\omega,T),
      \end{equation}
%70
with geometrical factors \cite{BTJ}
\begin{equation}
 \label{eq gf}
  L_{\|} = \frac{1}{(1+R_{\|}/R_{\bot})^{1.6}}\,, \qquad L_{\bot} = \frac{1}{2}(1-L_{\|})\,.
   \end{equation}
%71

Finally, it should be reminded that the real and imaginary parts of a dielectric constant
tensor $\epsilon_{jj}$ can be easily obtained with the help of tensor $\sigma_{jj}$, as~\cite{L}
\begin{align}
 \label{eq eps1}
&  \epsilon'_{jj}(\omega,T) = \epsilon_{\infty} - \frac{4\pi}{\omega}\sigma''_{jj}(\omega,T),\\
%    \end{equation}
%72
%\begin{equation}
 \label{eq eps2}
&  \epsilon''_{jj}(\omega,T) = \frac{4\pi}{\omega}\sigma'_{jj}(\omega,T).
    \end{align}
%73
The sign in the last equation and before the second term in equation~(\ref{eq eps1}) corresponds to the exponent sign
in time dependence of an external electric field ${\bf E}(t) = {\bf E}_0 \re^{-\ri\omega t}$.
The imaginary part of the dielectric constant provides basically the damping and broadening of SPR.

The linewidth of SPR and FOM are strongly dependent on the particle radius. We will illustrate this
dependence below, having built the temperature behavior of SPR linewidth for different $R$ as an example.

\begin{figure}[htb]
\centerline{\includegraphics[width=0.65\textwidth]{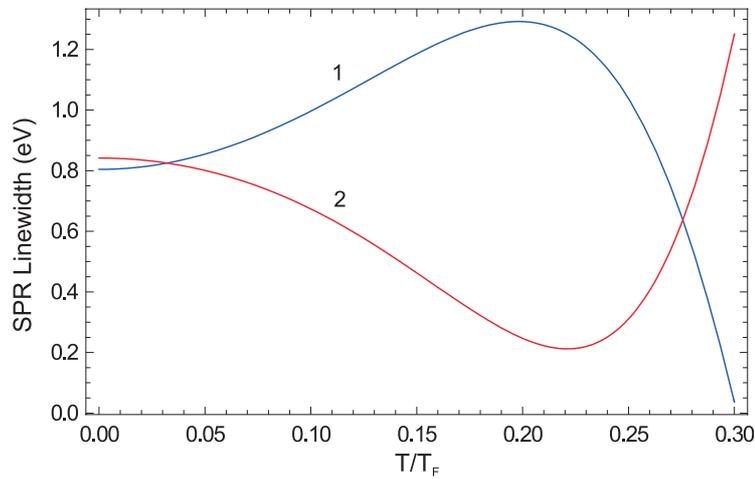}}
\caption{(Color online) SPR linewidth vs temperature for
spherical Au particles with different $R\; (\AA)$: $50$ (1) and $54.7$ (2), embedded in water.}
\label{fig-1}
\end{figure}

The SPR linewidth is plotted in figure~\ref{fig-1} against temperature for spherical Au nanoparticles with
two different radii embedded in the water ($n=1.33$). Calculations were made according to the
formulas (\ref{eq gam}) and (\ref{eq sig1}) for the following parameters of Au
nanoparticle~\cite{K}: $\upsilon_\textrm{F} = 1.394\times 10^{8}$~cm/s, $T_\textrm{F} = 6.41\times 10^{4}$~K,
$\omega_{\rm pl} = 1.37\times 10^{16}$~s$^{-1}$,
and $\epsilon_{\infty}=9.84$. The electron concentration was estimated from
equation~(\ref{eq_ect})
with $n_0$ given above by equation~(\ref{eq n}). For Au, $n_0\simeq 5.9\times 10^{22}$~cm$^{-3}$.
The behavior of $\Gamma(T)$ is governed by the temperature dependence of the $\epsilon''(T)$
in accordance with equations~(\ref{eq gam}) and (\ref{eq eps2}).

High temperature of an electron gas can be achieved inside the MN by using laser pumping
at SPR frequency when hot electrons are produced. As the temperature of an electron gas
decreases after an initial excitation, the SPR linewidth can increase (see curve 1, figure~\ref{fig-1})
or decrease (curve 2) depending on the particle radius. Such different behavior is due to
the linewidth oscillations with the particle radius changing~\cite{G2}.
Earlier~\cite{RMV,YBG}, the results for Ag and Au nanoparticles were reported where the linewidth
is only reduced linearly with the temperature lowering over a range of crystal temperatures.

Let us calculate the polarizability tensor for MNs at finite temperatures.

\section{Polarizability tensor}
\label{sec:3}

The dipole electric moment for a spherical particle embedded in the
media with dielectric constant $\epsilon_\textrm{m}$, can be written as follows:
\begin{equation}
 \label{eq dm}
  d_j(\omega) = \epsilon_\textrm{m}\sum_{k=1}^3\alpha_{jk}(\omega)\, E_{k}(\omega),
    \end{equation}
%74
where $E_{k}$ is the $k$ component of an external electric field
and $\alpha_{jk}$ is the polarizability tensor of the MN.
The calculation of ${\bf d}$ is especially simple for the case
when the particle sizes are small compared to some
``wavelength'' $\delta\sim c/(\sqrt{|\epsilon|}\;\omega)$,
which corresponds to the frequency $\omega$ in the particle
bulk. In this case, one can calculate the polarizability of MN using
the formulae obtained for an external uniform statical
field.
In general terms, this tensor can be treated as being a complex value
\begin{equation}
 \label{eq cma}
  \alpha_{jk}(\omega) = \alpha'_{jk}(\omega) + \ri \alpha''_{jk}(\omega).
    \end{equation}
%75

Tensor of  electric conductivity was introduced by equation~(\ref{eq ca}).
The component of the inner electric field in equation~(\ref{eq ca}) is linked
to the component of an external electric filed in equation~(\ref{eq dm})  by expression
(\ref{eq if}). Since the density of electric current is related to
the dipole moment by a simple relation
\begin{equation}
 \label{eq jd}
  j(t) = \frac{1}{V}\frac{\partial}{\partial t} d(t),
   \end{equation}
%76
it is easy to establish the connection between polarizability and conductivity
of MN. Performing the Fourier transformation of equation~(\ref{eq jd}), we obtain
\begin{equation}
 \label{eq jod}
  j(\omega) = \frac{\ri\omega}{V}\;d(\omega).
   \end{equation}
%77
Then, using equations~(\ref{eq dm}), (\ref{eq jod}),
and (\ref{eq ca}), the polarizability tensor can be expressed
through the conductivity tensor $\alpha_{jj}$ by means of
\begin{equation}
 \label{eq sial}
  \alpha_{jj}(\omega,T) = -\frac{V}{\epsilon_\textrm{m} \omega} \;
   \frac{\ri\,\sigma_{jj}(\omega,T)}{1+L_j[\epsilon_{jj}(\omega)/\epsilon_\textrm{m}-1]}\,.
    \end{equation}
%78
In equation~(\ref{eq sial}) for MN with a spheroidal
shape, parameter $V$ represents the volume of spheroid $(=\frac{4}{3}\pi R^2_{\bot} R_{\|})$.

Using equations~(\ref{eq eps1}) and (\ref{eq eps2}) for an isolated MN {\em in host},
we come to the Clausius-Mossotti dipole polarizability \cite{BH}
\begin{equation}
 \label{eq alep}
   \alpha_{jj}(\omega,T) = \frac{V}{4\pi L_j}\; \frac{\epsilon_{jj}(\omega,T)-\epsilon_\textrm{m}}
    {\epsilon_{jj}(\omega,T)+(1/L_j-1)\epsilon_\textrm{m}}\,,
     \end{equation}
%79
where tensor $\epsilon_{jj}$  corresponds to a given frequency $\omega$,
and in the case of frequencies close to the plasma oscillations of electrons
in metal, it can be presented as follows:
\begin{equation}
 \label{eq epsg}
  \epsilon_{jj}(\omega,T)\simeq \epsilon_{\infty}-\left(\frac{\omega_{\rm pl}}{\omega}
   \right)^2+\frac{4\pi}{\omega} \ri\sigma_{jj}(\omega,T).
    \end{equation}
%80
Optical absorption generally measures the ${\rm Im}\,\epsilon(\omega,T)$.

\section{Light absorption and scattering}
\label{sec:4}

In classical electrodynamics it is supposed that the light scattering results from
the polarization of the scattered particle when it is illuminated with a beam of light.
Since the spatially uniform field causes only a dipole polarization, the nanoparticle
scatters the light like  a vibration dipole~\cite{BH}.
In the case of particles of a {\em spheroidal shape}, the dipole
moment of the MN can be presented in the form~\cite{G3}
\begin{equation}
 \label{eq edp}
  {\bf d}=\alpha_{\bot} \epsilon_m {\bf E}+(\alpha_{\|}-\alpha_{\bot}) \epsilon_m({\bf m E}){\bf m}\,,
   \end{equation}
%81
where ${\bf m}$ is the unit vector directed along the spheroid axis of revolution.
Accounting for ${\bf m\cdot E} = E\cos\phi$, where $\phi$ is the angle between
the spheroid axis of revolution and the direction of the electric field ${\bf E}$,
the equation~(\ref{eq edp}) can be rewritten as follows:
\begin{equation}
 \label{eq sdp}
  {\bf d} = \sqrt{\alpha^2_{\bot}\sin^2\phi+\alpha_{\|}^2\cos^2\phi} \,\,\,\epsilon_m {\bf E}\,.
   \end{equation}
%82

Using our forerunning formulae for ${\bf d}$ and $\alpha$, it is easy to calculate
the temperature dependence of light absorption and scattering. The overall absorption
and scattering cross-sections for a spheroidal MN are given by
\begin{align}
 \label{eq acs}
  C_\textrm{abs} &= 4\pi\frac{\omega}{c}\sqrt{\epsilon_\textrm{m}}
   \left[({\rm Im}\,\alpha_{\|}) \cos^2\phi + ({\rm Im}\,\alpha_{\bot}) \sin^2\phi\right],\\
%    \end{equation}
%83
%\begin{equation}
 \label{eq scs}
  C_\textrm{sca}& = \frac{8}{3}\pi\left(\frac{\omega}{c}\sqrt{\epsilon_\textrm{m}}\right)^4
   \left(|\alpha_{\|}|^2\cos^2\phi + |\alpha_{\bot}|^2\sin^2\phi\right).
    \end{align}
%84
The square of diagonal components of the polarizability
tensor can be presented in the Lorenzian form as follows:
\begin{equation}
 \label{eq alpp}
   |\alpha_{\|\choose\bot}|^2=\left[\frac{V}{4\pi L_{\|\choose\bot}}\right]^2
    \frac{\left[(1-\xi_\textrm{m})\omega^2-\omega^2_{\|\choose\bot}\right]^2+
     \left[2\omega\gamma_{\|\choose\bot}\right]^2 }{\left[\omega^2-
      \omega^2_{\|\choose\bot}\right]^2+\left[2\omega\gamma_{\|\choose\bot}\right]^2}\,,
       \end{equation}
%85
and the imaginary part of the polarizability tensor is
\begin{equation}
 \label{eq imal}
   {\rm Im}\,\alpha_{\|\choose\bot}=\left[\frac{V}{4\pi L_{\|\choose\bot}}\right]
    \frac{2\omega^3\xi_\textrm{m}\gamma_{\|\choose\bot} }{\left[\omega^2-
      \omega^2_{\|\choose\bot}\right]^2+\left[2\omega\gamma_{\|\choose\bot}\right]^2}\,.
       \end{equation}
%86
Here,
\begin{equation}
 \label{eq xi}
  \xi_\textrm{m}=\frac{\epsilon_\textrm{m}}{\epsilon_\textrm{m}
  +L_{\|\choose\bot}(1-\epsilon_\textrm{m})}\,,
   \end{equation}
%87
\begin{equation}
 \label{eq omres}
  \omega^2_{\|\choose\bot} \equiv  \omega^2_{\|\choose\bot}(T) = \frac{L_{\|\choose\bot}}{
   \epsilon_\textrm{m}+L_{\|\choose\bot}(1-\epsilon_\textrm{m})} \;\omega^2_\textrm{pl}(T)
    \end{equation}
%88
are the $\|$ and $\bot$ surface plasmon frequencies, and
\begin{equation}
 \label{eq gamm}
  \gamma_{\|\choose\bot}\equiv\gamma_{\|\choose\bot}(\omega, T)=\frac{2\pi L_{\|\choose\bot}}
   {\epsilon_\textrm{m}+L_{\|\choose\bot}(1-\epsilon_\textrm{m})} {\rm Re}\, \sigma_{\|\choose\bot}(\omega, T)
    \end{equation}
%89
is the half-width of the resonance for light polarized along ($\|$)
or across ($\bot$) the rotation axis of the spheroid, $\sigma_{\|}$,
$\sigma_{\bot}$ are the corresponding components of the conductivity
tensor, and $L_{\|}$, $L_{\bot}$ are the geometrical factors
given above by equations~(\ref{eq gf}).

On resonance, the polarizability tensor looks as follows:
\begin{equation}
 \label{eq alr}
   \alpha_{\|\choose\bot}(\omega,T) \simeq \frac{V}{8\pi}
    \frac{\xi_\textrm{m}\omega_{\|\choose\bot}(T)}{L_{\|\choose\bot}
    \gamma_{\|\choose\bot}(\omega,T)}\,,
     \end{equation}
%90
with $\gamma_{\|\choose\bot}$ taken at $\omega=\omega_{\|\choose\bot}$.
The contribution of nonresonant frequencies is so small that
their input into the scattering cross section can be neglected.

In the case of MN having a spherical shape $L_{\|}=L_{\bot}=1/3$,
equations~(\ref{eq xi})--(\ref{eq gamm}) become as follows:
\begin{equation}
 \label{eq xisp}
  \xi_\textrm{m} = \frac{3\epsilon_\textrm{m}}{2\epsilon_\textrm{m}+1},
   \end{equation}
%91
\begin{equation}
 \label{eq omsp}
  \omega^2_{\|\choose\bot}(T) \equiv \omega_\textrm{sph} =
   \frac{\omega^2_\textrm{pl}(T)}{2\epsilon_\textrm{m}+1}\,,
    \end{equation}
%92
\begin{equation}
 \label{eq gasp}
  \gamma_{\|\choose\bot}(\omega,T) \equiv \gamma_\textrm{sph}(\omega,T) =
   \frac{2\pi}{2\epsilon_\textrm{m}+1}\,\sigma(\omega,T)\,.
    \end{equation}
%93
Then, the expressions (\ref{eq acs}) and (\ref{eq scs})
can be easily transformed to the forms
\begin{align}
 \label{eq absp}
   C_\textrm{abs}(\omega,T) &= 9 V \varepsilon_\textrm{m}^{3/2}\frac{\omega}{c}
    \frac{\epsilon''(\omega,T)} {\left[\epsilon'(\omega,T)+2\epsilon_\textrm{m}\right]^2+
     \left[\epsilon''(\omega,T)\right]^2}\,,\\
%      \end{equation}
%94
%\begin{equation}
 \label{eq scsp}
   C_\textrm{sca}(\omega,T) &= \frac{3V^2}{2\pi}\epsilon^2_\textrm{m} \left(\frac{\omega}{c}\right)^4
    \frac{[\epsilon'(\omega,T)-\epsilon_\textrm{m}]^2+[\epsilon''(\omega,T)]^2}{\left[
     \epsilon'(\omega,T)+2\epsilon_\textrm{m}\right]^2+\left[\epsilon''(\omega,T)\right]^2}\,,
      \end{align}
%95
where we have combined equations~(\ref{eq xisp})--(\ref{eq gasp}) with the real and imaginary
parts of equation~(\ref{eq epsg}). The most intense cross-sections are observed at the frequency
which corresponds to the plasmon resonance of a {\em spherical} nanoparticle in the vacuum.
Formulae (\ref{eq scsp}) and (\ref{eq absp})
yield a resonance when $\epsilon'(\omega,T)$ $ = -2\epsilon_\textrm{m}$.
The absorption and the scattering cross-sections defined by equations~(\ref{eq acs})
and (\ref{eq scs}) reach the maximal values at the frequencies of the SPRs
$\omega = \omega_{\|\choose\bot}$ as well. Near the surface, plasmon resonance light may
interact with the particle (at low temperatures) over a cross-sectional area usually
larger than the geometric cross section of the particle because the polarizability
of the particle becomes very high in this frequency range.

We calculate the scattering efficiency of the MN, which is defined as the ratio
\begin{equation}
 \label{eq efe}
  S_\textrm{ef}=\frac{C}{C_\textrm{geom}}\,, \qquad C_\textrm{geom} = \pi\left(\frac{c}{\omega_\textrm{pl}}\right)^2,
   \end{equation}
%96
where $C_\textrm{geom}$ is the geometrical cross-section of an individual particle
and $C$ is defined by equations~(\ref{eq acs}) and (\ref{eq scs}), or by (\ref{eq absp}) and (\ref{eq scsp}).

\begin{figure}[h]
\centerline{\includegraphics[width=0.65\textwidth]{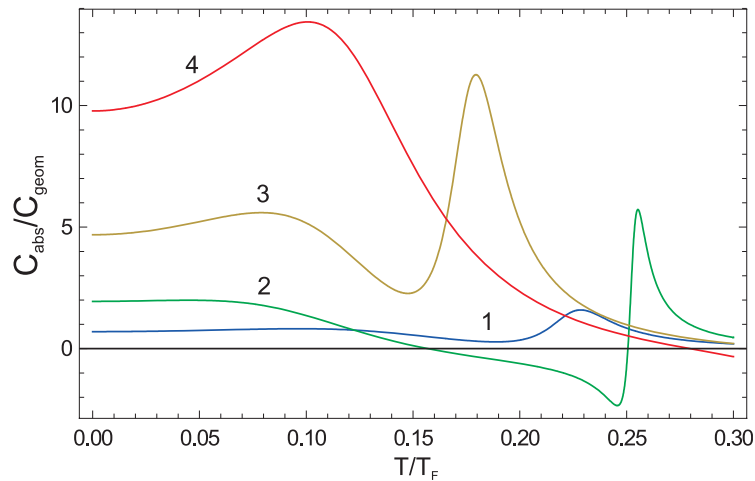}}
\caption{(Color online) The absorption
crossection vs temperature for spherical Au particles with different
$R\; (\AA)$: $78$ (1), $100$ (2), $125$ (3), and $150$ (4),
embedded in water.}
\label{fig_2}
\end{figure}
Figure~\ref{fig_2} shows the variation of an absorption efficiency as a function of temperature
for  gold nanoparticles having a spherical shape embedded in water ($n=1.33$). The calculations
were performed in  accordance with equations~(\ref{eq efe}) and (\ref{eq absp}), with the use of the
same parameters as above for figure~\ref{fig-1}. For zero temperature, we get the classical result~\cite{BH,KV},
when the SPR intensity depends on the particle radius: larger particles have a larger scattering
or absorption cross sections. The situation  drastically changes at finite electron temperatures.
As we can see from figure~\ref{fig_2}, the absorption efficiency not only increases (right hand sides of
curves 1--4) with the temperature drop, but can also decrease (left hand sides of curves 1--4).
We have an interesting situation when the absorption crossection with the temperature lowering
increases at first, reaches its peak, then decreases, and starts to increase again.
These observations are explained by the conductivity of MN, which oscillate
with the particle radius changing~\cite{G}.

\begin{figure}[htb]
\centerline{\includegraphics[width=0.65\textwidth]{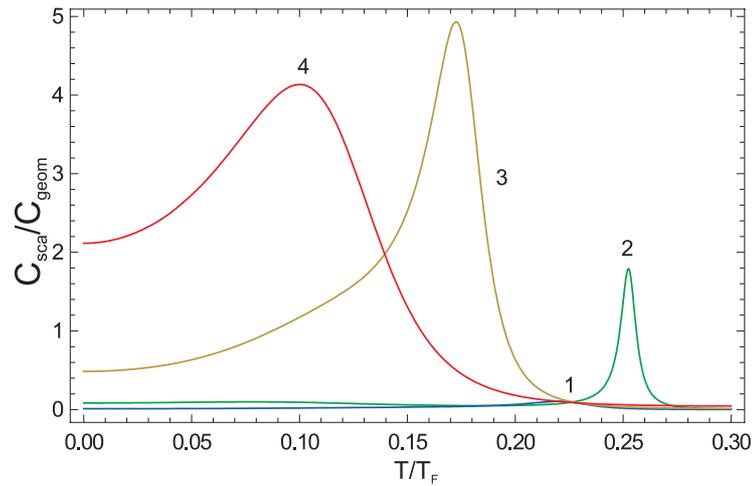}}
\caption{(Color online)
The same as in figure~\ref{fig_2}, but for scattering crossection.}
\label{fig-3}
\end{figure}

In the case of light scattering (see figure~\ref{fig-3}), the effect of MN size is more pronounced
because the scattering cross-section is proportional to the square of the particle volume.

In figure~\ref{fig-3}, one can see that the efficiency of light scattering at a resonance frequency does not only increase (curves 1--3) with the growth of the MN radius, but can also decrease
(curve 4). The latter means that the efficiency of scattering at a resonance frequency
can be suppressed with the temperature rise for some particle radii.
 Depending on the electron temperature, the scattering efficiency reaches the maximum value,
 whose peak position depends on the nanoparticle radius.

The peak positions of the resonance plasmon absorption (or scattering) by Au nanoparticle
embedded in water can be shifted toward shorter wavelengths as the temperature of electron
gas is lowered after an initial excitation. This behavior is consistent with the known
temperature-dependent shift of the differential transmission peak in an isolated sodium
nanoparticle~\cite{WWI}.

\vspace{2ex}
\section{Discussion of results}
\label{sec:5}

Here, we shortly discuss the effects caused by deviations from the spherical particle shape.
When the symmetry of the particle decreases, the number of resonance peaks increases.
If the shape of a particle deviates from the spherical one, the 3-fold degeneracy dipole mode splits
into two (for spheroid) or three (for ellipsoid) modes giving rise to corresponding scattering
peaks. In this case, the plasmon response strongly depends on the particle position relative to the
illumination direction. To evaluate the electron temperature effect on the absorption or scattering
crossections for various illumination directions, one can perform numerical calculations
using equations~(\ref{eq acs})--(\ref{eq gamm}) and (\ref{eq sigt}).

The particle absorption or scattering effects depend on the size and shape of  individual
particles. Metallic particles that are much smaller than the wavelength of light tend
to absorb still more and hence the absorption dominates. The scattered light intensity from small
MNs is extremely low. Indeed, Rayleigh scattering theory predicts that the scattering cross
section of a spherical cluster should drop as the sixth power of the diameter.

The quantity that is measured experimentally is frequ\-ency-resolved time-dependent
transmission~\cite{WWI} $\Upsilon_\textrm{on}$ after the excitation of the MN by a pump laser pulse.
Then, the static equilibrium transmission $\Upsilon_\textrm{off}$ of the sample measured in the
absence of the pump is subtracted to obtain the relative differential transmission~\cite{BMC}
$\delta\Upsilon/\Upsilon =$$(\Upsilon_\textrm{on}-\Upsilon_\textrm{off})/\Upsilon_\textrm{off}$.
In the thermal equilibrium, the $\delta\Upsilon$ can be related to the temperature
dependent absorption cross section by~\cite{HBH}
\begin{equation}
 \label{eq ef}
  \frac{\delta\Upsilon}{\Upsilon} = -\frac{3}{2\pi R^2}\left[C_\textrm{abs}(\omega,T)-C_\textrm{abs}(\omega,T_0)\right],
   \end{equation}
where $T_0$ is the temperature of the MN before excitation.
This relation holds for very small sizes of MN when the reflectivity of the MN is very small.

A direct comparison of theoretical results with most of the available experimental
measurements of the optical properties of MNs is still a matter of debate because
inhomogeneities  in nanoparticle size, shape, and local environment hide the homogeneous
width of the surface plasmon resonance.

\section{Conclusions}
\label{sec:6}
	
We have applied a kinetic theory to the calculation of the optical properties of a metal
nanoparticle embedded in a dielectric media at different temperatures.
It permits to determine the cross-section of light absorption or scattering for various
polarizations of the incident electromagnetic wave. The obtained analytical formulas
provide the evaluation of the dynamics of light absorption or scattering intensities
at the plasmon frequencies for nanoparticles with different radii and shapes, when
the temperature of electron gas becomes settled after the action of a laser pulse.

The analytical expressions for electroconductivity and polarizability tensors have been
obtained for the case of MNs with a spherical shape. They permit to calculate a number of other physical quantities (e.g., resonance linewidth, figure of merit, etc.) for a particular temperature.
We theoretically studied  the variation of both the linewidth of SPR and the efficiency
of the light absorption and scattering by MN with the change of the electron temperature.
The case where the frequency of a laser beam is close to surface plasmon
frequencies of a spheroidal MN is studied in detail.

 The high sensitivity of the temperature dependence of SPR linewidth to the radius
 of a particle was established. Even a small variation in the particle radius
can drastically change the run of the SPR linewidth curve vs temperature.
This is due to the linewidth oscillations when the radius of MN is varied.
The efficiencies of light absorption and scattering for various temperatures
strongly depend on the particle radius as well. The Au nanoparticles
with different radii are used for illustration.

\section*{Acknowledgement}
The author is indebted to Prof. P.M.~Tomchuk for helpful discussions and useful comments.

%
%% If you have problems with typesetting in ukrainian uncomment lines below.
%
%  \lastpage
%  \end{document}

\newpage
\ukrainianpart

\title{Температурна залежність плазмонних резонансів у сфероїдальних металевих наночастинках}
\author{М.І. Григорчук}
\address{
Інститут теоретичної фізики ім.~М.М.~Боголюбова НАН України, \\
 вул.~Метрологічна, 14--б, 03680 Київ, Україна
}

\makeukrtitle

\begin{abstract}
\tolerance=3000%
В рамках кінетичної теорії вивчається вплив температури електронів на поглинання
та розсіювання світ\-ла металевими наночастинками (МН) при збудженні електронних
коливань поверхневого плазмона. Одержані формули для тензорів електропровідності
та поляризованості для кінцевих температур електронного газу. Детально вивчені
електрична провідність та напівширина резонансу поверхневого плазмона для сферичної МН.
Досліджується ефективність поглинання та розсіювання світла при зміні температури
в залежності від розмірів МН. Зокрема, знайдено, що ефективність поглинання може
не тільки збільшуватися при зменшенні температури, а також й зменшуватись.
Одержані формули дозволяють аналітично обчислювати різноманітні оптичні й транспортні
явища для МН довільної сфероїдальної форми, що знаходяться в довільному діелектричному середовищі.
\keywords електронна температура, металеві наночастинки, електропровідність, тензор поляризованості,
резонанс поверхневого плазмона

\end{abstract}

\end{document}